\documentclass[journal=nalefd,manuscript=article]{achemso}

\usepackage{graphicx,bm}
\usepackage{float}
\usepackage{amsmath}
\usepackage{amssymb}
\usepackage{color}

\newlength\figurewidth
\author{Taegeun Song}
\email{tsong@ictp.it}
\affiliation[The Abdus Salam International Centre for Theoretical Physics]
{The Abdus Salam International Centre for Theoretical Physics, Trieste, Italy}

\author{Mikhail N. Kiselev}
\affiliation[The Abdus Salam International Centre for Theoretical Physics]
{The Abdus Salam International Centre for Theoretical Physics, Trieste, Italy}

\author{Konstantin Kikoin}
\affiliation[Tel-Aviv University]
{School of Physics and Astronomy, Tel-Aviv University, Tel-Aviv, Israel}

\author{Robert I. Shekhter}
\affiliation[University of Gothenburg]
{Department of Physics, University of Gothenburg, G\"oteborg, Sweden}

\author{Leonid Y. Gorelik}
\affiliation[Chalmers University of Technology]
{Department of Applied Physics, Chalmers University of Technology, G\"oteborg, Sweden}

\title[\texttt{achemso} demonstration]
{\color{black} Self-sustained oscillations in nanoelectromechanical systems induced by Kondo resonance}
\color{black}

\begin{document}
\begin{abstract}
 We investigate instability and dynamical properties of  nanoelectromechanical systems
represented by a single-electron device containing movable quantum dot
attached to a vibrating cantilever via asymmetric tunnel contact.
{\color{black}The Kondo resonance in electron tunneling between source and shuttle facilitates self-sustained oscillations originated from strong coupling of mechanical and electronic/spin degrees of freedom.
We analyze stability diagram for two-channel Kondo shuttling regime due to limitations given by the electromotive force acting on a moving shuttle and
find that the saturation amplitude of oscillation is associated with the retardation effect of Kondo-cloud.
The results shed light on possible ways of experimental realization of dynamical probe for the Kondo-cloud
by using high tunability of mechanical dissipation as well as supersensitive detection of mechanical displacement}.
\end{abstract}

The recent progress in nano-technology made it possible to fabricate
nano-devices in which mechanical degrees of freedom are strongly coupled
not only to electronic {\color{black}charge} [nanoelecromechanics (NEM)] but also
to spin degrees of freedom [nanospintromechanics (NSM)]\cite{shuttlereview,kis06}.
While manipulating {\color{black}the charge} degrees of freedom requires the energies/external voltages determined by the Coulomb
interaction in the nano-device (charging energy of quantum dot), the spin manipulation needs much smaller
scales of the energy  determined by the exchange interaction. Therefore, on the one hand, the spin manipulation is free of a heating problem, and, on the other hand, {\color{black}it allows one to achieve} very high efficiency of devices\cite{prlkiselev}.

A special case where the spin degrees of freedom are dominant in quantum transport is the Kondo-effect which manifests itself as a resonance scattering of electrons on the impurity spin.\cite{kondo,glazman} The retardation effects in NEM devices
result in two-channel Kondo tunneling which enhances both spin and charge transport
due to maximal overlap of the wave functions of the electrons in the leads \cite{kis06, prlkiselev}. 
These processes are mediated by the spin flip of the localized state in the dot.
One more facet of the Kondo effect is formation of
a screening cloud of conduction electrons which "dresses" the spin of quantum impurity.
The typical length scale of screening cloud is $\sim 1\mu m$\cite{bor}.
There has been several proposals to detect the size of the Kondo  cloud\cite{prlkiselev,jinhong,guber,holzner,affleck,affleck2,mitchell,bosser}, however,
the unambiguous results are still unavailable  and
there has been no conclusive measurement due to the quantum fluctuations with zero averaged spin\cite{boy}.

We are interested in new effects where a moving quantum impurity is nano-machined by attaching it to a nano-mechanical device. Such devices are realized as quantum dots incorporated in a mechanical system
which oscillates between two metallic leads.
These mechanical systems include long cantilever nanorelay \cite{Kin03}, atomic force microscope with a tip \cite{AFM1, AFM2}
and nanoisland attached either to the cantilever \cite{Azuma07} or to one of the leads \cite{AFM1, AFM2}.
The basic understanding of the NEM/NSM has been achieved in  both theoretical
\cite{gorelik,Sheht05,Sheht09,Shekht06,Shekht10,Shekht11}
and experimental\cite{Hut09,Weig12} studies of single electron shuttling.
Alternatively, the mechanical system
can play itself a role of one of the contacts when the quantum dot (impurity)
is deposited on a top of metallic cantilever.{\color{black} \cite{Azuma07}} In these cases either one or two
tunnel barriers change its shape in the process of mechanical motion thus providing
a coupling of mechanical and electronic/spin degrees of freedom. The temporal dynamics of the Kondo cloud
is governed by two main effects. First, the cloud adiabatically follows a position of quantum impurity. 
Second, the size of the cloud changes in time due to change of tunnel matrix elements.
Both these effect are accompanied by the retardation processes similar to those which
determine the polaronic effects due to strong electron-phonon interaction.
How does the dynamics of the Kondo cloud affects the mechanical system?
How can one probe this dynamics? Is it possible to control the cloud's size?
Some of these questions have been addressed in a recent publication.\cite{prlkiselev}. This study has shown that the mechanical dissipation is controlled by the kinetics of Kondo screening if
electric dc current is transmitted through the system in the presence of
external magnetic field. Besides, the characteristic time
determining the kinetics of Kondo screening may be measured through
the mechanical quality factor. Thus, the strong coupling of spin with mechanical subsystem
allows a superhigh tunability of mechanical dissipation as well as supersensitive
detection of mechanical displacement.
\medskip

In this paper we address a question of whether such a strong coupling between
mechanics and spintronics can drive system from nearly adiabatic regime of small amplitude mechanical vibrations to a steady state regime with large amplitude self-sustained oscillations.
As an example of such regime we consider an instability associated with
appearance of self-sustained oscillations in the system induced by "Kondo friction". It will be shown that this regime can be controlled by
electric (source-drain voltage, gate voltage) and magnetic fields.
We analyse the sensitivity of solutions to the initial conditions and
construct the complete phase diagram of the model. We show that the system
possesses reach non-linear dynamics (Hopf-pitchfork bifurcation).
We demonstrate that by controlling the displacements (velocities) of mechanical system with a high precision
one can manipulate both spin and charge tunnel currents.

 A sketch of the system under consideration is presented  in Fig.1. A nano-island is mounted on the metallic cantilever attached to the drain electrode. The distance between the source electrode  and the island, and thus the tunnel coupling  between them, depends on the cantilever motion. An external magnetic field is applied perpendicular to the cantilever far from island.
In our consideration the flexural vibration of the nanowire is restricted to the dynamics of the fundamental mode only. It is treated as a damped harmonic oscillator with frequency $\omega_0$, and  quality factor $Q_{0}$.

\begin{figure}
\includegraphics[width=\figurewidth]{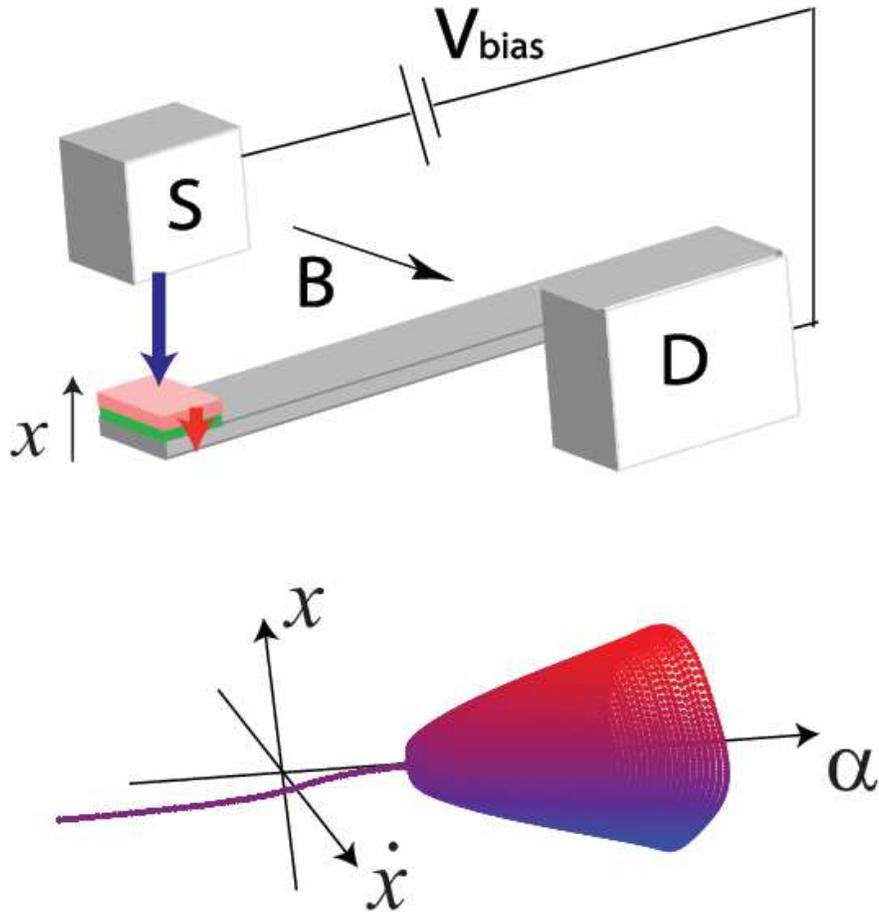}
\vspace{5mm}
 \caption{(Upper panel) Shuttle with a cantilever: the quantum dot (pink) is deposited on top of a metallic cantilever. The constant width tunnel barrier between the dot and metallic reservoir is depicted by green color. The source-shuttle barrier changes its width when the device oscillates.  Blue and red arrows indicate an asymmetry of tunnel barrier widths. The device is subject to external source (S) - drain (D) voltage $V_{\rm bias}$ and external magnetic field $B$ applied perpendicular to the plane of oscillations.
$x$ denotes a dimensionless displacement 
(in units of tunnel length) measured from the equilibrium position of the shuttle.
(Lower panel) The evolution of the phase space  $\{x,\dot{x}\}$ as a function of universal parameter
$\alpha$ (dimensionless force, see below for discussion).
One can see the appearance of instability at some
critical value of this control parameter.
 }
 \label{fig:f1}
 \end{figure}

The equation of mechanical motion for Kondo-NEM coupling device is given by:
\begin{eqnarray}
\ddot{u}+\frac{\omega_{0}}{Q_{0}}\dot{u}+\omega_{0}^{2}u
=\frac{1}{m}\left[(L\cdot I \cdot B)+ F_K\right] \label{eqofmotion}
\end{eqnarray}
where $u$ describes the cantilever's displacement of free end (see Fig.1) and $m$ is effective mass.
The right hand side of Eq. (\ref{fig:f1}) includes the Lorentz force acting on a metallic cantilever in the presence of {\color{black} the effective magnetic field $B$ and the "Kondo-force" associated with the coordinate dependence of the ground state Kondo energy $E_{gs} \sim T_K(u)$ \cite{Hewson,com1}. Here $L$ is the cantilever length and {\color{black} 
the current $I$ through the cantilever equals to $I=I_{dc}+I_{ac} + I_{emf}$.
Two first terms  contributing to the tunnel current have been calculated in the adiabatic approximation in the limit of strong Kondo coupling at $T\ll T_K$ where the Kondo temperature $T_K$ depends parametrically on time {\color{black} via cantilever vibration} in a reasonable assumption that the mechanical motion of a shuttle is slow in comparison with all characteristic times for Kondo tunneling: \cite{prlkiselev}}
$I_{dc}$ corresponds to a time-modulated dc component, $I_{ac}$ \color{black} is an ac component
associated with the modulation of the Kondo cloud deeply inside the leads.
\begin{eqnarray}\label{ac}
I_{dc}&=& 2 G_{0} V_{bias}\cosh^{-2}\left[\frac{u(t)-u_{0}}{\lambda}\right]  \\ \label{dc}
I_{ac}&=&\frac{\dot{u}(t)}{\lambda} \frac{eE_{c}}{8\Gamma_{0}}\frac{eV_{bias}}{k_{B}T_{K}(t)}
\frac{\tanh(\frac{u(t)-u_{0}}{\lambda})}{\cosh^{2}(\frac{u(t)-u_{0}}{\lambda})}
\end{eqnarray}
Here $G_{0}$=$e^2/h$ is the unitary conductance \color{black} per spin projection,} \color{black} $V_{bias}$ is bias voltage, $\lambda$ is the tunneling length for the source-island tunnel barrier.
The Kondo temperature for moving island is
$$
k_BT_K(t)\equiv k_B T_K[u(t)]=D_0\exp\left[-\frac{\pi E_c}{4(\Gamma_s(u) + \Gamma_d)}\right],
$$
where $E_c$ is the charging energy of the dot, 
$\Gamma_{d}=\Gamma_{0}$ and $\Gamma_s(u)=\Gamma_{0}\exp\left\{2(u-u_{0})/\lambda\right\}$ are the island-drain (d) and island-source (s) tunnel rates, $D_0$ is effective bandwidth for the electrons in the leads.
\color{black}
Thus, $I_{dc}$ describes the Ohmic regime where the time dependence is associated
with $\cosh^{-2}[(u(t)-u_0)/\lambda]$ Breit-Wigner factor given by the time-dependent tunnel widths. In contrast to it, the major time dependence of $I_{ac}$ is connected with  time modulations of the Kondo temperature $T_K(t)$.

{\color{black}The last contribution to the current $I$  is $I_{emf}=-G_{0} \dot{u} L B$. This term is related to \color{black} the voltage difference
\color{black} $\mathcal{E}=-\dot{u}LB$ between the electrodes induced by motion of the metallic cantilever in the effective magnetic field $B$.
As a result, the velocity dependent current term $I_{ac}$ is modified by the factor $1-\Delta(u)$, where $\Delta\propto (\Gamma_{0}/E_{c})(k_{B}T_{K}/eV_{bias})(\lambda LB/\phi_{0})$,  $\phi_{0}=h/(2e)$ is a magnetic flux quanta. Thus, the electromotive force \color{black} $\sim B^2$ \color{black} is immaterial in the regime of weak magnetic fields.

\color{black}First we analyse the amplitude dynamics  and stability of the system without emf term, and then consider a regime where the emf term plays an important role.
It is convenient to introduce the dimensionless equation of motions using Eq. (\ref{eqofmotion})
scaled by tunnel length $\lambda$, ($x\equiv u(t)/\lambda$) and dimensionless time  scaled with $\omega_{0}^{-1}$:
\begin{eqnarray}
\ddot{x}+\gamma \dot{x}+x= \frac{\alpha_K {\color{black}(t)}+\alpha}{\cosh^{2}(x-x_{0})}-\dot{x} \alpha \tau_{\beta}f(x-x_{0}), \label{eomdim}
\end{eqnarray}
In these notations $\alpha=\frac{2 G_{0}V_{bias}BL}{m\omega_{0}^{2}\lambda}$, $\alpha_K {\color{black}(t)}
=-\frac{\pi E_c {\color{black} k_{B} T_{K}(t)}}{{\color{black}8} \Gamma_0 m\omega_0^2\lambda^2}$ are the dimensionless Lorentz and Kondo forces respectively,
$\beta=\frac{\pi}{4} \frac{E_{c}}{ \Gamma_{0}}$ is the coupling strength of electronic states,
$\gamma=\frac{1}{Q_{0}}$ is the mechanical friction coefficient
and the $x_{0}$ is dimensionless parameter $x_{0}=u_{0}/\lambda$ characterizing the asymmetry of the system
at the equilibrium position such that \color{black}$\Gamma_l(x_0)/\Gamma_0\neq 1$\color{black}. The retardation time associated with dynamics of the Kondo cloud $\tau_{\beta}=\frac{\hbar \omega_{0}}{k_{B}T_{K}^{min}}\frac{\beta}{2}$ is parametrically large compared to the time of Kondo cloud formation \cite{prlkiselev}.
The correction to the quality factor {\color{black}$Q_0$} of mechanical system due to retardation effects is determined by functional form of $f(x)=\frac{\tanh(x)}{\cosh^2(x)}e^{-\frac{\beta}{2}(1+\tanh(x))}$.
The time dependent Kondo temperature in the strong coupling limit at $T\ll T_{K}^{min}$ is given by
\begin{eqnarray}
k_{B}T_{k}(t) = k_{B}T_{K}^{min}\exp\left\{\frac{\beta}{2}[1+\tanh(x(t)-x_{0})])\right\}.
\end{eqnarray}
The $k_{B}T_{K}^{min}$ plays the role of the cutoff energy for Kondo problem.
{\color{black}As is mentioned above,} we consider adiabatically slow motion of the QD,  $\hbar\omega_{0}\ll k_{B}T_{K}^{min}$
provided the condition $\{k_{B}T, g\mu_{B}B,eV_{bias}\} \ll k_{B}T_{K}^{min}$ is fulfilled.

In order to analyse the amplitude dynamics in the regime of high-quality resonator $\frac{k_{B}T_{K}}{\hbar\omega_{0}}\ll Q_{0}$
we apply the Krylov-Bogoliubov averaging method\cite{kbmethod}.
The equations for amplitude dynamics can be obtained by means of the ansatz $x(t)=A(t)\sin(\omega_{0}t+\phi)$. In this approximation we ignore the dynamics of the phase $\phi$.
The equation for amplitude dynamics for Eq. (\ref{eomdim}) is written as:
\begin{eqnarray} \label{amplitude}
\dot{A} &=& -\frac{\gamma}{2}A-\frac{\alpha\tau_{\beta}}{2 \pi}A\int_{0}^{2\pi}\cos^{2} \theta f (A\sin \theta-x_{0}) d\theta.
\end{eqnarray}

The results of numerical analysis of Eq. (\ref{eomdim}) are shown in Fig. \ref{fig:amp} A.
{\color{black}At zero bias} ($\alpha=0$), Eq. (\ref{eomdim}) describes a
damped harmonic oscillator with the friction $\gamma$. In this case the system is characterized by single stable attracting fixed point at the origin (black line in Fig.\ref{fig:amp} A.)

When the finite bias is applied to the system in the presence of magnetic field perpendicular to the plane,
{\color{black}the increase of the Lorentz force results in the change} of behaviour of the oscillator at some critical value
$\alpha=\alpha_{c}$. As a result, the equation for the amplitude (\ref{amplitude}) acquires an
additional non-trivial fixed point for $\dot{A}=0$  indicated by the red line in Fig. \ref{fig:amp} A.  {\color{black}In the regime of $\alpha<\alpha_{c}$ with single attracting fixed point at $A=0$ an
assumption $x\approx x_{0}$ still can be adopted}. Under this approximation Eq. (\ref{amplitude}) can be solved analytically. Eq. (\ref{eomdim}) can also be solved by applying the Taylor expansion of hyperbolic functions. Equivalently, the system can be treated as a damped harmonic oscillator with effective friction coefficient
$\gamma_{eff}=\gamma(1-\frac{\alpha\tau_{\beta}x_{0}}{\gamma}e^{-\frac{\beta}{2}})$.
In the regime $\alpha\gtrsim\alpha_{c}$, the system is characterized by two attracting and one repelling fixed points in the space of parameters $\{\alpha,x_{0}\}$ {\color{black}determined by equilibrium position of the shuttle, bias and magnetic field as control parameters.}
In this regime the system shows bi-stability and flows either to the fixed point at the origin corresponding to damped oscillations
or to the regime of self-sustained oscillation depending on the initial conditions.
At $\alpha\gg\alpha_{c}$, the system falls to the self-sustained oscillations regime.

In Fig.\ref{fig:amp} B we plot a saturation amplitude of the system as a function of $\alpha$.
The hysteresis of the system is originated from the coexistence of two fixed points characterizing a damped  and self-sustained oscillation in the intermediate regime.
Moreover, there exists a regime of linearly increasing saturation amplitude.
Approximating $\tanh{x}=x$, for $|x|<1$, and $\tanh{x}={\rm sign}[x]$ for $|x| >1$, we rewrite the condition for $\dot{A}=0$  as
$$\frac{1}{\pi}\int_{-1}^{1}\frac{1}{A}\sqrt{1-\left(\frac{\xi+x_{0}}{A}\right)^2}\xi\left(1-\xi^{2}\right)e^{-\frac{\beta}{2}\xi}d\xi=-\frac{\gamma}{\alpha\tau_{\beta}}e^{\frac{\beta}{2}},$$
where $\xi\equiv A \sin\theta-x_{0}$.
As a result, the saturation amplitude is found as $A_{sat}=\frac{8}{\pi}\frac{1}{\gamma\beta}\frac{\hbar\omega_{0}}{k_{B}T_{K}^{min}}\alpha$,
giving rise to the linear slope
($A_{sat}\propto Q_{0}\frac{\hbar\omega_{0}}{k_{B}T_{K}^{min}}\frac{\Gamma_{0}}{E_{c}}$).

\vspace{5mm}
\begin{figure}
\includegraphics[width=\figurewidth]{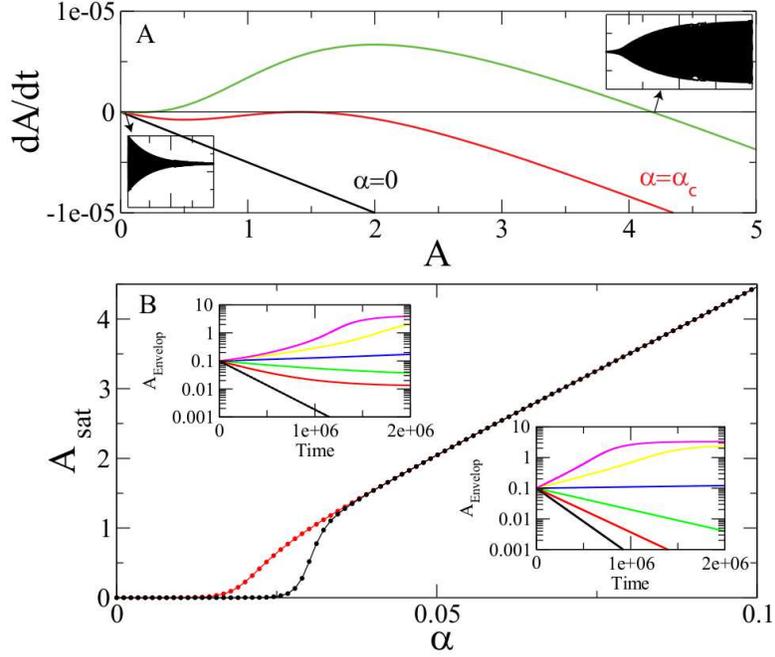}
\vspace{5mm}
 \caption{{\color{black} Panel A:} Amplitude dynamics at different values of the dimensionless force $\alpha$
  (see details in the text). Insets: time trace of the oscillation at two different fixed point indicated by arrow.
   {\color{black} Panel B}: Saturation amplitude as a function of dimensionless force. Different colors denote initial conditions near (black dots) and far (red dots) from the equilibrium position $x_{0}$.
Insets: {\color{black}amplitude envelope} as a function of dimensionless time calculated by using 
the last (velocity dependent) term in equation (\ref{eomdim}) responsible for the amplitude dynamics
(upper inset) and approximate equation (\ref{amplitude}) (lower inset).
The parameter $\alpha$ varies from $\alpha=0$ (black) to $\alpha=0.1$ (magenta).
The equations are solved for the following set of parameters: $\beta=8$, $\gamma=10^{-5}$, {\color{black}$x_{0}=0.5$} and $\frac{\hbar\omega_{0}}{k_{B}T_{K}^{min}}=10^{-3}$.}
 \label{fig:amp}
\end{figure}

Next we analyse a stability of the system by linearizing the equations in the vicinity of the stable fixed point describing the stationary states.
It is convenient to rewrite Eq.(\ref{eomdim}) in equivalent form of
two coupled first order differential equations:
\begin{eqnarray}\label{firstorder}
\dot{x} &=& y \\  \dot{y} &=&
-x +\frac{{\color{black} \alpha_{K}(t)}+\alpha}{\cosh^{2}(x-x_{0})} \nonumber
-(\alpha \tau_{\beta}f(x) +\gamma) y
\end{eqnarray}
{\color{black}
While position of the fixed point $x^{*}$  can be found from the condition
$x^{*}\cosh^{2}(x^{*}-x_{0})=\alpha(1+\delta \exp[\frac{\beta}{2}(1+\tanh(x^{*}-x_{0}))])$,
the corresponding Jacobian matrix is given by
}
\[
\begin{array}{lc}
    J=\left(\begin{array}{@{}ccc@{}}
                    0 & 1  \\
                    {\color{black}
                    \frac{\alpha g(x^{*}-x_{0})}{\cosh^{2}(x^{*}-x_{0})}-1 }&
                    -\alpha\tau_{\beta}f(x^{*}-x_{0})-\gamma
                  \end{array}\right),
\end{array}
\]
{\color{black}
where the $g(x)=\delta(\frac{\beta}{2 \cosh(x)}-2\tanh(x))\exp[\frac{\beta}{2}(1+\tanh(x))]-2\tanh(x)$ and
 the $\delta$ is the ratio between $\alpha$ and Kondo force at the minimal Kondo temperature. }
Interestingly, this condition allows a regime of multiple solution for $x^{*}$ depending on $|x_{0}|$ and $\alpha$.
In the Fig.\ref{fig:stab}  we plot the stability diagram of the Jacobian matrix in the parameter space {\color{black}$\{\alpha,x_{0}\}$.}

\vspace{5mm}
\begin{figure}
\includegraphics[width=\figurewidth]{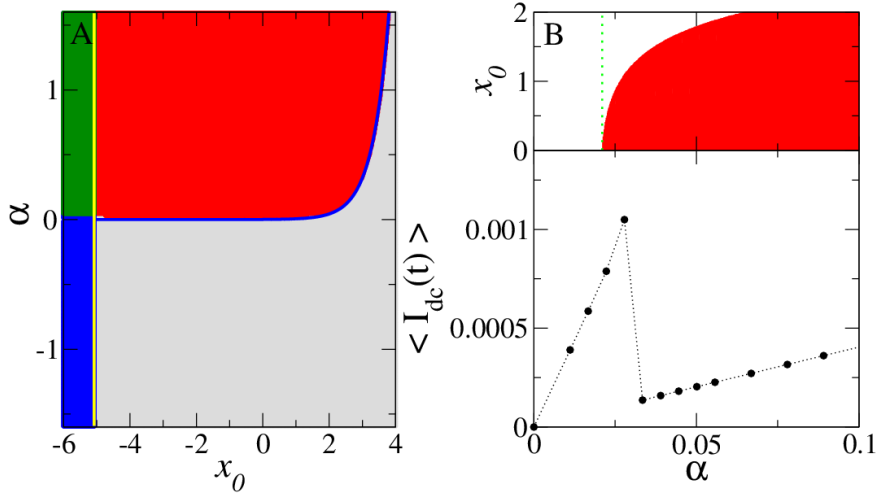}
\vspace{5mm}
 \caption{
{\color{black} A:} Stability diagram in the phase space of asymmetry parameter $x_{0}$ and dimensionless force $\alpha$.
Different colors correspond to different classes of stability: stable focus (gray regime), unstable focus (red and green regimes depending on initial conditions), and saddle point (green and blue regimes, see the text for details).
The weak out of equilibrium condition used in calculations  is determined by
{\color{black} |$\frac{e V_{bias}}{k_{B}T_{K}^{\min}}\cdot\frac{g\mu_{B}B}{k_{B}T_{K}^{\min}}|\leq0.1$, and $\alpha\cdot\delta=50.27$}.
 The blue solid lines are drawn in accordance with  Eq.(\ref{codition1}). 
 \color{black} Yellow solid line determines the boundaries of a multy-valued solution. \color{black} {\color{black} B: Upper panel is a fragment of phase diagram (A). The green dotted line represents
 a condition  $\alpha=\gamma/\tau_{\beta}|f({\color{black}x_{un}})|$ determined by the eigenvalues of Jacobian matrix. Lower panel shows the average current $\langle I_{dc}\rangle$ as a function of $\alpha$
 at $x_0=1.0$.}
 {\color{black} It is seen that the transition from unstable focus to stable focus can be realized by changing the direction of magnetic field $B \to -B$ at given $V_{bias}$.}}
 \label{fig:stab}
\end{figure}

The linearized system can be categorized by a stable focus, unstable focus, and a saddle point.
Analysing conditions for each category we obtain the eigenvalues of the Jacobian matrix $\lambda_{\pm}$
{\color{black}
\begin{eqnarray}\nonumber
\lambda_{\pm}&=&-\frac{1}{2} (\alpha\tau_{\beta}f(x^{*}-x_{0})+\gamma) \\ \nonumber &\pm&
\frac{1}{2} \sqrt{(\alpha\tau_{\beta}f(x^{*}-x_{0})+\gamma)^{2} - 4 \left( 1- \frac{\alpha g(x^{*}-x_{0})}{\cosh^{2}(x^{*}-x_{0})} \right)}
\end{eqnarray}
}

First, we consider a regime of single solution for the fixed point $x^{*}$.
In this case instability arises in the absence of stable focus (negative quality factor characterizing an increment of the oscillations). {\color{black} The negative $Q$ corresponds to pumping regime, where the system is effectively "heated" in contrast with damping regime of $Q>0$, which may be interpreted as effective cooling of a shuttling device.}
The positive values of ${\rm Re}(\lambda_{\pm})$ give rise to the regime of instability when
{\color{black} both $\alpha > 0$ and $\alpha > \frac{\gamma}{\tau_{\beta}|f(x_{un})|}$ are satisfied}.
The critical values for the instability are given by a functional shape of $f(x)$ with
$\alpha=\gamma/\tau_{\beta}|f({\color{black}x_{un}})|$, (green dotted line in Fig.\ref{fig:stab} B),
{\color{black} where
 $x_{un}\cong x_{0}+\tanh^{-1}{(\frac{3-\sqrt{9+\beta(\beta+2)})}{\beta})}\approx x_{0}+\tanh^{-1}(-1+2/\beta)$.
Thus, the critical values of $\alpha$ for the unstable regime depending on the quality factor which is given by,
$\alpha_{+}^{un}\gtrsim k_{B}T_{K}/\hbar\omega_{0}\cdot1/Q_{0}$.
}
Second, the regime of multi-valued solution is determined by
{\color{black} $1<\frac{\alpha g(x^{*}-x_{0})}{\cosh^{2}(x^{*}-x_{0})}$
(yellow solid line in Fig.\ref{fig:stab} A)
}
The saddle point solution leads to a bi-stability of the system under certain condition for $Q$ factor.

The approximate solution determining the boundaries for the instability regime  of applied magnetic field is given by:
{\color{black}
\begin{eqnarray}
\frac{ \gamma}{4 \tau_{\beta}}\exp\left[-\frac{\beta}{2}+2 x_{0}\right]<\alpha,
\label{codition1}
\end{eqnarray}
}
{\color{black} This condition is valid for the range of magnetic fields;
$
\frac{m\omega_{0}^{2}\lambda}{eV_{bias}}
\frac{\phi_{0}}{L}
\frac{k_{B}T_{K}}{\hbar\omega_{0}}
\frac{1}{Q_{0}}
<~B.
$
}
The upper limit of this domain of validity is determined by the smallest value of two contributions, namely the emf force and the asymmetry condition.
Taking into account all necessary constraints for the stability regimes we construct the phase diagram of our model (see Fig. \ref{fig:diagram}).
This phase diagram shows the boundaries for the self-sustained oscillations regime (gray). Two green dotted lines correspond to two different values of the parameter $\alpha$, while the green arrow goes in a direction of increasing $\alpha$. The red line is defined by the condition $B<\frac{E_{x}}{\Gamma_{0}}\frac{eV_{bias}}{k_{B}T_{K}^{min}}\frac{\lambda}{L}\frac{\phi_{0}}{\lambda^{2}}$
{\color{black} (see \cite{prlkiselev})}.
The black line is determined by the critical value of asymmetry parameter
$x_c$ as a functions of $Q$-factor and Kondo temperature $T_K^{min}$: $x_{c}=\frac{k_{B}T_{K}^{min}}{\hbar\omega_{0}}\frac{1}{Q_{0}}$.
Thus, the upper limit of instability regime under  condition of fixed asymmetry parameter $x_{c}<x_{0}$ is described by
{\color{black}
$B/B_0<\frac{E_{c}}{\Gamma_{0}}\frac{eV_{bias}}{k_{B}T_{K}^{min}},$
}
where $B_0$ is magnetic field corresponding to the flux quanta through $\lambda\cdot L$ square. {\color{black} The periodic splashes of $I_{dc}$ and $I_{ac}$ near the turning points of the shuttle\cite{prlkiselev} are shown in the lower and upper insets, respectively. The change of average dc current $\langle I_{dc}(t)\rangle$ (\ref{dc}) at transition from damping oscillations regime to self-sustained oscillation regime with increasing $\alpha$ is illustrated in Fig. 3B.}

\vspace{5mm}

\begin{figure}
\includegraphics[width=\figurewidth]{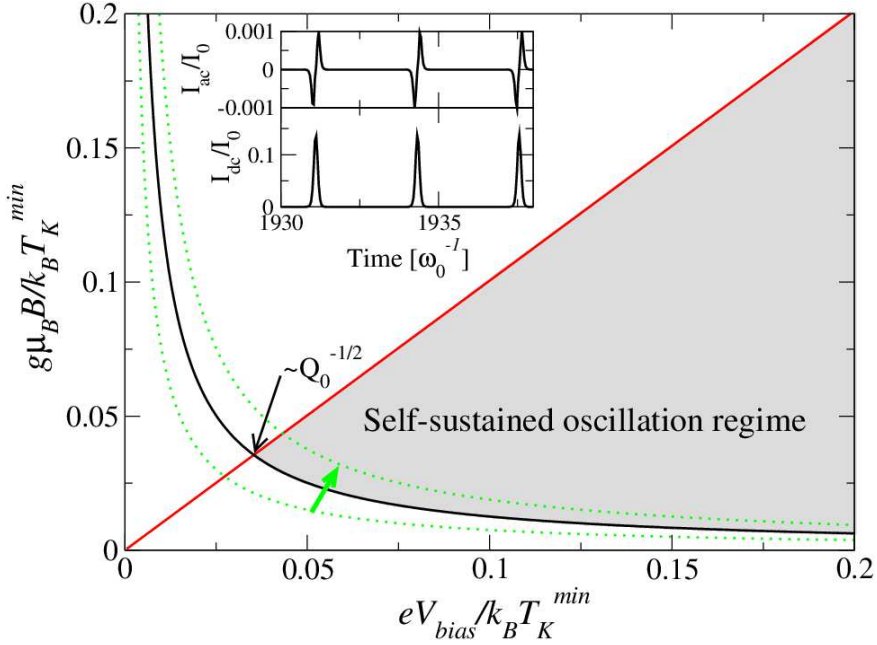}
\vspace{5mm}
 \caption{Stability (phase) diagram in a parameter space of $\frac{g\mu_{B}B}{k_{B}T_{K}}$, and $\frac{eV_{bias}}{k_{B}T_{K}}$.
 The calculations are performed at
 {\color{black} $\beta=8$, $\frac{\hbar\omega_{0}}{k_{B}T_{K}^{min}}=10^{-3}$,  $\alpha\cdot\delta=50.25$.
 The red line is defined by the condition $B<\frac{E_{x}}{\Gamma_{0}}\frac{eV_{bias}}{k_{B}T_{K}^{min}}\frac{\lambda}{L}\frac{\phi_{0}}{\lambda^{2}}$.
 {\color{black} 
The parameter $\alpha$ increases for transition from lower to upper green dotted line (see details in the text). Green arrow shows how one enters the self-sustained oscillation regime by changing the force at given anisotropy parameter.}
 Inset: time trace of tunnel current contribution in steady state regime under the parameters of
 $ x_0=0.5$, $ \frac{eV_{bias}}{k_{B}T_{k}^{min}}=0.1$,  $ \frac{g\mu_{B}B}{k_{B}T_{k}^{min}}=0.1$, and $I_{0}=\frac{m\omega_{0}^{2}\lambda}{BL}$. }
 }
 \label{fig:diagram}
\end{figure}

\color{black}
Plugging in some typical values of $m$, $\lambda$, $\omega_0$ and $Q_0$ into the conditions for upper and lower bounds of the instability regime: $~m=10^{-19} - 10^{-21}~kg,~ \lambda=1\mathring{A}
$, and $\omega_{0}=10^{7} - 10^{9} Hz$, $Q_0 = 10^{5} - 10^{7}$
allows one to estimate the range of magnetic fields and bias voltages
for the self-sustained oscillations in Kondo shuttling regime: $1~T <B<10~T$, $\frac{eV_{bias}}{k_{B}T_{K}^{min}}=0.1,~\frac{\hbar\omega_{0}}{k_{B}T_{K}^{min}}=10^{-3}$. This range of parameters covers
cantilever materials from light single-wall carbon nanotubes to relatively heavy SiN. The range of
quality factors refers to best known nano-mechanical devices \cite{Weig13}.
\color{black}
\medskip

Summarizing, we analysed a full fledged stability diagram of the Kondo shuttle device subject both
to variation of external dc electric and magnetic fields and asymmetry of the tunnel barriers. {\color{black}Kondo effect with its anomalously long relaxation time of dynamical spin screening is an ideal tool for coupling spin and mechanical degrees of freedom.}
We have shown that the competition between the mechanical damping of the oscillator at zero field, zero bias and contribution coming from the strong resonance spin scattering (Kondo effect) results {\color{black}in the loss of mechanical stability manifested in two different regimes of NEM/NSM oscillations}. Namely,
if the Kondo force controlled by external fields further damps the oscillator,  we obtain an efficient mechanism of cooling the nano-shuttle. On the other hand, if the contribution of the Kondo force
enhances the oscillations, we arrive at the {\color{black} non-linear steady state regime of self sustained oscillations}. We have found the critical values of external fields
and asymmetry parameter determining the {\color{black}instability regimes for adiabatic Kondo shuttling.} The phase diagram of the
Kondo shuttle model is constructed by taking into account the limitations given by the electromotive force
which always contributes to the friction. We have shown that due to exponential sensitivity of Kondo effect to external parameters and strong coupling between the mechanical and electron (spin) degrees of freedom, {\color{black}the device including an element with localized spin acquires super-high tunability}. We suggested {\color{black}the experimental setup to realize the Kondo shuttling instability and estimated all necessary conditions for it. We believe that experiments with such setup can provide valuable
 information on kinetics of formation of the Kondo cloud, and eventually the Kondo shuttle can be used for experimental
spin manipulation in nano-spintromechanical devices.}

\acknowledgement

We appreciate illuminating discussions with B.L. Altshuler, Ya. Blanter, S. Flach, F. Marquardt, N. Prokof'ev and E. Weig. The research of MNK was supported in part by the National Science Foundation under Grant No. NSF PHY11-25915. The research of KK was partially supported by ISF Grant No. 400/12. The work of RIS and LYG was supported in part by Swedish VR.


\end{document}